\documentclass[sn-mathphys-num]{sn-jnl}
\usepackage{graphicx}%
\usepackage{multirow}%
\usepackage{amsmath,amssymb,amsfonts}%
\usepackage{amsthm}%
\usepackage{mathrsfs}%
\usepackage[title]{appendix}%
\usepackage{xcolor}%
\usepackage{textcomp}%
\usepackage{manyfoot}%
\usepackage{booktabs}%
\usepackage{algorithm}%
\usepackage{algorithmicx}%
\usepackage{algpseudocode}%
\usepackage{listings}%
\usepackage{subcaption} 
\usepackage{tikz}
\usepackage{soul}
\usepackage{hyperref}
\usepackage{titlesec}

\renewcommand{\eqref}[1]{Equation~(\ref{eq:#1})}

\newcommand{\tabref}[1]{Table~\ref{tab:#1}}
\newcommand{\figref}[1]{Figure~\ref{fig:#1}}

\renewcommand{\Bar}{\overline}

\renewcommand{\vec}[1]{\ensuremath{\boldsymbol{#1}}}

\newcommand{\tvec}[1]{\ensuremath{\tilde{\boldsymbol{#1}}}}

\newcommand{\bvec}[1]{\ensuremath{\Bar{\boldsymbol{#1}}}}

\theoremstyle{thmstyleone}%
\theoremstyle{thmstyletwo}%

\theoremstyle{thmstylethree}%

\begin{document}

\title[Article Title]{Accelerated Real-time Cine and Flow Under In-magnet Staged Exercise}


\author[1]{\fnm{Preethi} \sur{Chandrasekaran}}\email{chandrasekaran.71@osu.edu}
\author[1]{\fnm{Chong} \sur{Chen}}\email{chong.chen@osumc.edu}
\author[2]{\fnm{Yingmin} \sur{Liu}}\email{yingmin.liu@osumc.edu}
\author[3]{\fnm{Syed Murtaza} \sur{Arshad}}\email{arshad.32@osu.edu}
\author[4]{\fnm{Christopher} \sur{Crabtree}}\email{crabtree.223@osu.edu}
\author[5]{\fnm{Matthew} \sur{Tong}}\email{matthew.tong@osumc.edu}
\author[5]{\fnm{Yuchi} \sur{Han}}\email{yuchi.han@osumc.edu}
\author*[1,3]{\fnm{Rizwan} \sur{Ahmad}}\email{ahmad.46@osu.edu}

\affil[1]{\orgdiv{Biomedical Engineering}, \orgname{Ohio State University}, \orgaddress{ \city{Columbus}, \state{OH}, \country{USA}}}
\affil[2]{\orgdiv{Davis Heart and Lung Research Institute}, \orgname{Ohio State University Wexner Medical Center}, \orgaddress{ \city{Columbus}, \state{OH}, \country{USA}}}
\affil[3]{\orgdiv{Electrical and Computer Engineering}, \orgname{Ohio State University}, \orgaddress{ \city{Columbus}, \state{OH}, \country{USA}}}
\affil[4]{\orgdiv{Kinesiology}, \orgname{Ohio State University}, \orgaddress{ \city{Columbus}, \state{OH}, \country{USA}}}
\affil[5]{\orgdiv{Cardiovascular Medicine}, \orgname{Ohio State University Wexner Medical Center}, \orgaddress{ \city{Columbus}, \state{OH}, \country{USA}}}

\abstract{
\textbf{Background:} Cardiovascular magnetic resonance imaging (CMR) is a well-established imaging tool for diagnosing and managing cardiac conditions. The integration of exercise stress with CMR (ExCMR) can enhance its diagnostic capacity. Despite recent advances in CMR technology, quantitative ExCMR during exercise remains technically challenging due to motion artifacts and limited spatial and temporal resolution.

\textbf{Methods:} This study investigated the feasibility of biventricular functional and hemodynamic assessment using real-time (RT) ExCMR during a staged exercise protocol in 24 healthy volunteers. We employed high acceleration rates and applied a coil reweighting technique to minimize motion blurring and artifacts. We further applied a beat-selection technique that identified beats from the end-expiratory phase to minimize the impact of respiration-induced through-plane motion on cardiac function quantification. Additionally, results from six patients were presented to demonstrate clinical feasibility.

\textbf{Results:} Our findings indicated a consistent decrease in end-systolic volume and stable end-diastolic volume across exercise intensities, leading to increased stroke volume and ejection fraction. The selection of end-expiratory beats modestly enhanced the repeatability of cardiac function parameters, as shown by scan-rescan tests in nine volunteers. High scores from a blinded image quality assessment indicated that coil reweighting effectively minimized motion artifacts.

\textbf{Conclusions:} This study demonstrated the feasibility of RT ExCMR with in-magnet exercise in healthy subjects and patients. Our results indicate that high acceleration rates, coil reweighting, and selection of respiratory phase-specific heartbeats enhance image quality and repeatability of quantitative RT ExCMR.
}




\maketitle

\subsubsection*{List of Abbreviations}
AAo, ascending aorta; AP, arbitrary respiratory phase; APHR, age-predicted maximal heart rate; BBP, bean bag positioner; BMI, body mass index; BSA, body surface area; bSSFP, balanced steady-state free-precession; BRPE, Borg rating of perceived exertion; bpm, beats per minute; CAVA, Cartesian sampling with variable density and adjustable temporal resolution; CCC, concordance correlation coefficient; CMR, cardiovascular magnetic resonance imaging; CO, cardiac output; CR, coil reweighting; CS, compressed sensing; EDV, end-diastolic volume; ECG, electrocardiogram; EE, end-expiratory phase; EF, ejection fraction; ESV, end-systolic volume; ExCMR, exercise stress CMR; GRO, golden ratio offset; GRE, gradient echo; HR, heart rate; LV, left ventricle; MAD, mean absolute difference; MPA, main pulmonary artery; NFF, net forward flow; NMAE, normalized mean absolute error; R, acceleration rate; RT, real-time; RV, right ventricle; SCoRe, sparsity adaptive composite recovery; SV, stroke volume; UWT, undecimated wavelet transform; VENC, velocity encoding gradient; Vmax, peak velocity.

\section{Introduction}\label{sec:int}
Cardiovascular magnetic resonance imaging (CMR) is an established tool with proven diagnostic and prognostic value. It is considered a gold standard for biventricular volume quantification and detection of myocardial scar \cite{lima2004cardiovascular}. When paired with pharmacological or exercise stress, CMR-based first-pass perfusion can accurately diagnose coronary artery disease \cite{de2012diagnostic, arai2023stress} by identifying myocardial ischemia \cite{sakuma2005diagnostic}. More recently, cardiac stress imaging has emerged as an attractive option to investigate conditions other than ischemic heart disease. For example, myocardial contractile reserve under stress is observed to be a key prognostic factor in cardiomyopathies and heart failure and is independently associated with major cardiovascular events \cite{indorkar2019global,kobayashi2008dobutamine}. 

Stress CMR investigations commonly involve the use of vasodilators or inotropic agents \cite{kramer2020standardized}, which fail to mimic the hemodynamic alterations or provide the prognostic data associated with physical exercise \cite{leischik2007prognostic,pagnanelli2017pharmacologic}. CMR with exercise stress (ExCMR) has the ability to correlate symptoms (e.g., shortness of breath) and functional capacity with workload and thus offer a more comprehensive and dynamic assessment of cardiovascular function. In pulmonary artery hypertension, for example, ExCMR has shown a superior ability to capture dynamic changes in the right heart than any other modality \cite{goransson2019exercise}.

Several studies have demonstrated the use of CMR immediately after in-room treadmill exercise \cite{foster2012mr}. In multiple clinical trials, this strategy has demonstrated success in detecting myocardial ischemia \cite{raman2016diagnostic}. However, MRI-compatible treadmills are not readily available. Moreover, the need to transfer the patient from the treadmill to the MRI scanner not only creates a time delay between peak exercise and imaging but also makes imaging at various stress levels impractical. In contrast, the supine ergometers that enable exercise on the MRI patient table are more broadly available. Several studies have shown the feasibility of imaging during a pause in the in-magnet exercise \cite{le2016assessing, le2020application} or out-of-bore exercise using ergometers \cite{morales2022inline}. For the latter, the subject exercises on the MRI table outside the magnet. Once the target heart rate or exertion level is achieved, the subject is quickly moved into the scanner for imaging. Although more efficient than the treadmill protocol, these approaches still suffer from some of the same limitations.   


With advances in highly accelerated real-time (RT) imaging, ExCMR during in-magnet exercise has become increasingly feasible. In 2009, Lurz et al. demonstrated the feasibility of accelerated RT ExCMR in 12 healthy subjects with an isotropic spatial resolution of 3 mm \cite{lurz2009feasibility}. More recently, several other studies have shown the feasibility of RT ExCMR in small cohorts of healthy subjects \cite{beaudry2018exercise, craven2021exercise, li2021real} or patients \cite{edlund2022validation, backhaus2023cardiovascular}. Collectively, these studies and the technical advances described within highlight the ongoing evolution of ExCMR. However, these studies still suffer from one or more of these limitations: (i) lower acceleration rates leading to poor spatial or temporal resolution, (ii) long scan times, (iii) smaller number of slices leading to limited coverage, (iv) dependence on electrocardiogram (ECG) which is typically not reliable during exercise, (v) lack of right ventricular quantification, (vi) absence of hemodynamic assessment, (vii) offline image reconstruction, (viii) lack of demonstration for scan-rescan reproducibility, (ix) insufficient suppression of motion artifacts, especially under high-intensity exercise, (x) lack of a staged protocol, and (xi) sensitivity to through-plane motion.

In this study, we demonstrate the feasibility of biventricular hemodynamic and functional assessment of the heart using RT cine and flow data collected during a staged exercise protocol from 24 healthy subjects. More importantly, we present an in-magnet ExCMR protocol and related data acquisition and processing techniques, including coil reweighting to suppress motion artifacts, that address many of the technical limitations identified in recent ExCMR studies. To underline the clinical feasibility of RT ExCMR, we also present results from six patients.


\section{Methods}\label{sec:met}
\subsection{Data acquisition}
\begin{figure}[ht]
    \centering
    \includegraphics[width=\textwidth]{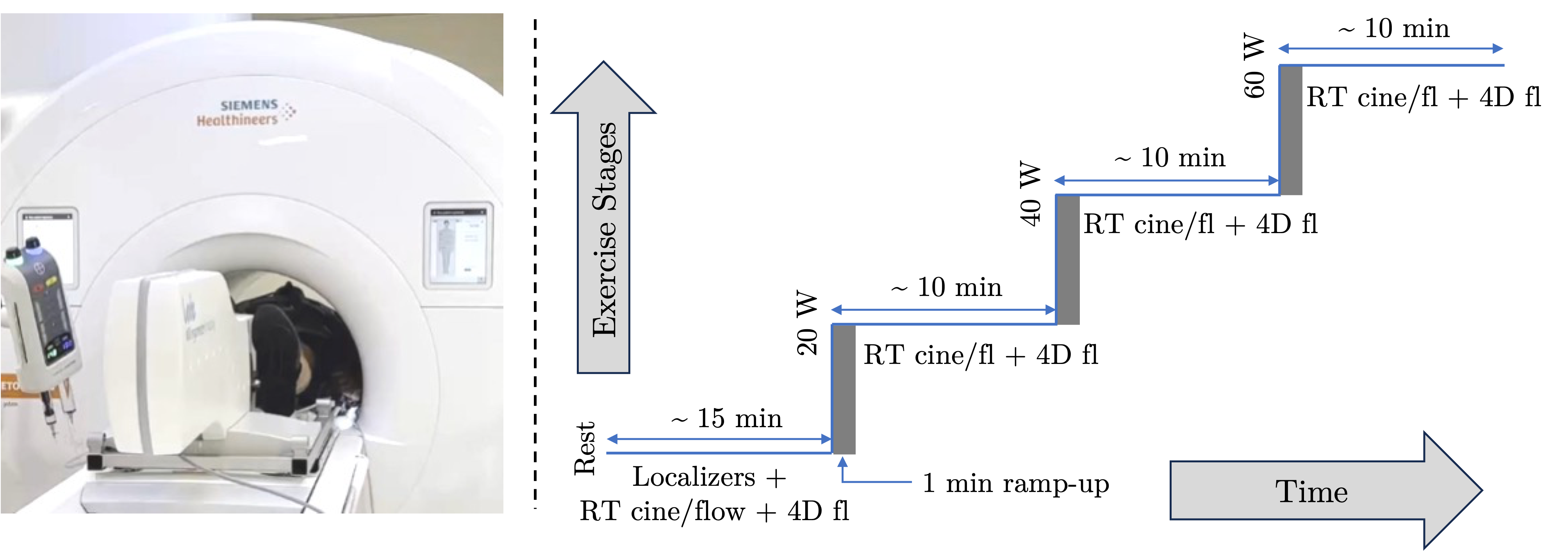}
    \caption{A volunteer exercising inside the MRI bore (left). The multi-stage ExCMR protocol used to scan healthy subjects (right).}
    \label{fig:protocol}
\end{figure}

Twenty-eight healthy volunteers over the age of 18 were recruited to participate in an institutional review board-approved ExCMR study. Written informed consent was obtained from each subject. Three volunteers were excluded owing to their non-compliance with instructions or incidental findings. As summarized in \tabref{demo}, twenty-five volunteers 
were imaged on a 3T scanner (MAGNETOM Vida, Siemens Healthcare, Erlangen, Germany) fitted with an in-magnet supine ergometer (MR Ergometer Pedal, Lode, The Netherlands). The imaging protocol included a free-breathing RT short-axis cine stack (11 to 14 slices) covering the whole heart and at least two long-axis cine slices. Additionally, phase-contrast CMR data were collected from two sets of three closely spaced slices for measuring RT flow at the roots of the aortic and pulmonic arteries, as well as from 4D flow imaging with whole-heart coverage. The RT cine data were collected using a balanced steady-state free-precession (bSSFP) sequence, while the data for RT flow and 4D flow were collected using a gradient echo (GRE) sequence. For cine, a pseudo-random Cartesian sampling pattern, called golden ratio offset (GRO) \cite{joshi2022technical}, was used for prospective undersampling, while a different Cartesian sampling pattern, called Cartesian sampling with variable density and adjustable temporal resolution (CAVA) \cite{rich2020cartesian}, was used for RT flow acquisition. GRO avoids large jumps in k-space between two consecutive readouts. This feature is important for avoiding eddy-currents-induced artifacts in bSSFP sequences \cite{bieri2005analysis}. In contrast, CAVA does not restrict the size of jumps in k-space but offers the added flexibility of changing the temporal resolution retrospectively. Since GRE sequences are less sensitive to large jumps in k-space, CAVA was used for RT flow. To maintain adequate spatial and temporal resolutions, both RT cine and flow acquisitions were highly accelerated, with acceleration rates of $R\geq 8$ and $R \geq 16$, respectively. The exercise protocol used to collect data is shown in \figref{protocol}, and the imaging parameters for RT cine and flow are summarized in \tabref{parameters}.


\begin{table}[ht]
\centering
\begin{tabular}{@{}lll@{}}
\toprule
                              & \textbf{Healthy Subjects}          & \textbf{Patients}       \\ \midrule
Number                        & $25$                      & $6$                    \\
Age (years)                   & $28.9\pm7.5$              & $60.5\pm14.2$            \\
Sex (M/F)                     & $16/9$                    & $3/3$               \\
BMI (kg/m\textsuperscript{2}) & $24.6\pm4.1$              & $31.3\pm8.0$             \\
BSA (m\textsuperscript{2})    & $1.4\pm0.2$               & $1.9\pm0.2$             \\ 
\bottomrule
\end{tabular}
\caption{Human subjects characteristics.}
\label{tab:demo}
\end{table}

\begin{table}[ht]
\centering
\begin{tabular}{@{}lll@{}}
\toprule
\textbf{Parameter/Sequence}   & \textbf{RT Cine}          & \textbf{RT Flow}       \\ \midrule
Sequence                      & bSSFP                     & GRE                    \\
Acquisition time (s/slice)    & $3 - 6$                   & $3 - 6$               \\
Acceleration rate             & $8 - 9$                   & $16 - 19$             \\
Repetition Time (ms)          & $2.60 - 3.16$             & $3.58$                   \\
Echo Time (ms)                & $1.10 - 1.42$             & $2.13$                   \\
Spatial Resolution (mm\textsuperscript{2})  & $(1.67 - 2.40) \times (1.67 - 2.40)$  & $(2.0 - 2.5) \times (2.0 - 2.5)$ \\
Temporal Resolution (ms)      & $34.8 - 41.6$             & $35.8 - 42.9$       \\
Flip Angle (deg)              & $24 - 45$                 & $10$                     \\
Slice thickness (mm)          & $6$                       & $6$                      \\
Sampling pattern              & Cartesian \cite{joshi2022technical}                 & Cartesian \cite{rich2020cartesian}             \\ 
Velocity encoding gradient (cm/s)                   & N/A                       & $100 - 150$ (rest), $200 - 300$ (stress)        \\ \bottomrule
\end{tabular}
\caption{MRI acquisition parameters for RT cine and RT flow sequences.}
\label{tab:parameters}
\end{table}

After the ergometer was attached to the end of the table, each subject was positioned on the scanning table, with a medium-sized bean bag positioner (BBP) under their upper torso. The subjects were instructed to find a comfortable position where they could pedal the ergometer without their knees touching the bore or over-extending their legs. Then, their feet were secured to the ergometer pedals using Velcro straps and suction was applied to BBP using a vacuum device to create a firm mold around the upper torso. This mold helped maintain the subject's position during in-magnet exercise. Finally, a blood pressure cuff, a finger pulse oximeter, and ECG leads were attached to the subjects to monitor their vital signs during the scan.

After data acquisition at rest, the volunteers were asked to pedal the ergometer at a rate of 60 cycles per minute by matching their cadence to a metronome played through the headphones. The initial workload was set to $20$ W and was increased in increments of $20$ W to the maximum of $60$ W. The same set of sequences was executed at each exercise stage after allowing a ramp-up time of one minute for the heart rate (HR) to stabilize. Each exercise stage lasted for 8 to 10 minutes. Due to exhaustion or leg fatigue, not all subjects were able to complete all three exercise stages. In total, we scanned $n=25$, $n=24$, $n=23$, and $n=16$ volunteers at rest, $20$ W, $40$ W, and $60$ W, respectively. In one of the volunteers, an incidental finding prompted the termination of the experiment after the resting stage. The data from this volunteer was only used in the repeatability study. 
For a subset of volunteers, we repeated RT cine and flow acquisitions in quick succession (within 2 minutes) at rest ($n = 11$) and at the $40$ W exercise stage ($n = 9$) to assess repeatability. The order of acquisition was RT cine, RT flow, RT cine (repeat), RT flow (repeat), and 4D flow. This study is focused on the analysis of RT cine and flow images; the data from the 4D flow acquisition will be reported separately \cite{arshad2024motion}.  

Before starting the exercise, all the volunteers were verbally encouraged to follow the metronome and maintain a steady cadence of 60 cycles per minute. If the cadence was observed to fall well below 60 cycles per minute during the exercise, the volunteers were offered words of encouragement to maintain the cadence to avoid increased resistance at a lower rotational frequency. 
Upon completing the scan, volunteers were requested to assess their level of exertion during each stage of the exercise using the Borg rating of perceived exertion (BRPE) scale \cite{williams2017borg}. This scale categorizes exertion into four distinct levels: ``no to extremely light exertion" (scores $6 – 8$), ``light exertion" (scores $9 – 12$), ``somewhat hard to hard" (scores $13 – 16$), and ``very hard to maximal exertion" (scores $17 – 20$). It is important to note that the scale's range from $6$ to $20$, when multiplied by 10, corresponds closely with HR measurements, underscoring its effectiveness in quantifying exertion levels. 

After completing the healthy subject study, six patient volunteers 
were also imaged on the same 3T scanner; see \tabref{demo}. Four patients had a history of unexplained dyspnea on exertion, and two patients had asymptomatic moderate valvular heart disease (one moderate aortic stenosis and one moderate mitral regurgitation). In contrast to the healthy subjects, a $10$ W workload increment was employed whenever the $20$ W increase was considered excessively challenging by the patient. The maximum exercise intensity achieved by the six patients was $60$ W, $30$ W, $40$ W, $20$ W, $30$ W, and $20$ W.

\subsection{Coil reweighting}

Utilizing the BBP significantly limited the torso movement. In addition, the subjects were instructed to firmly hold onto the handles connected to the table's rails during the exercise. Despite these measures, periodic bulk motion was observed during paddling in almost all cases. For some subjects, this resulted in significant degradation of image quality due to the physical movement of some of the receive coils resulting in temporally varying sensitivity maps. We attempted to use dynamic sensitivity maps, estimated from a sliding window of a small number of frames, but this approach was unsuccessful. Specifically, a smaller window (less than 8 frames) did not yield a fully sampled region of sufficient size or quality for reliable estimation of coil sensitivity maps, while a larger window introduced artifacts similar to those from time-averaged sensitivity maps.

\begin{figure}[ht]
    \centering
    \includegraphics[width=\textwidth]{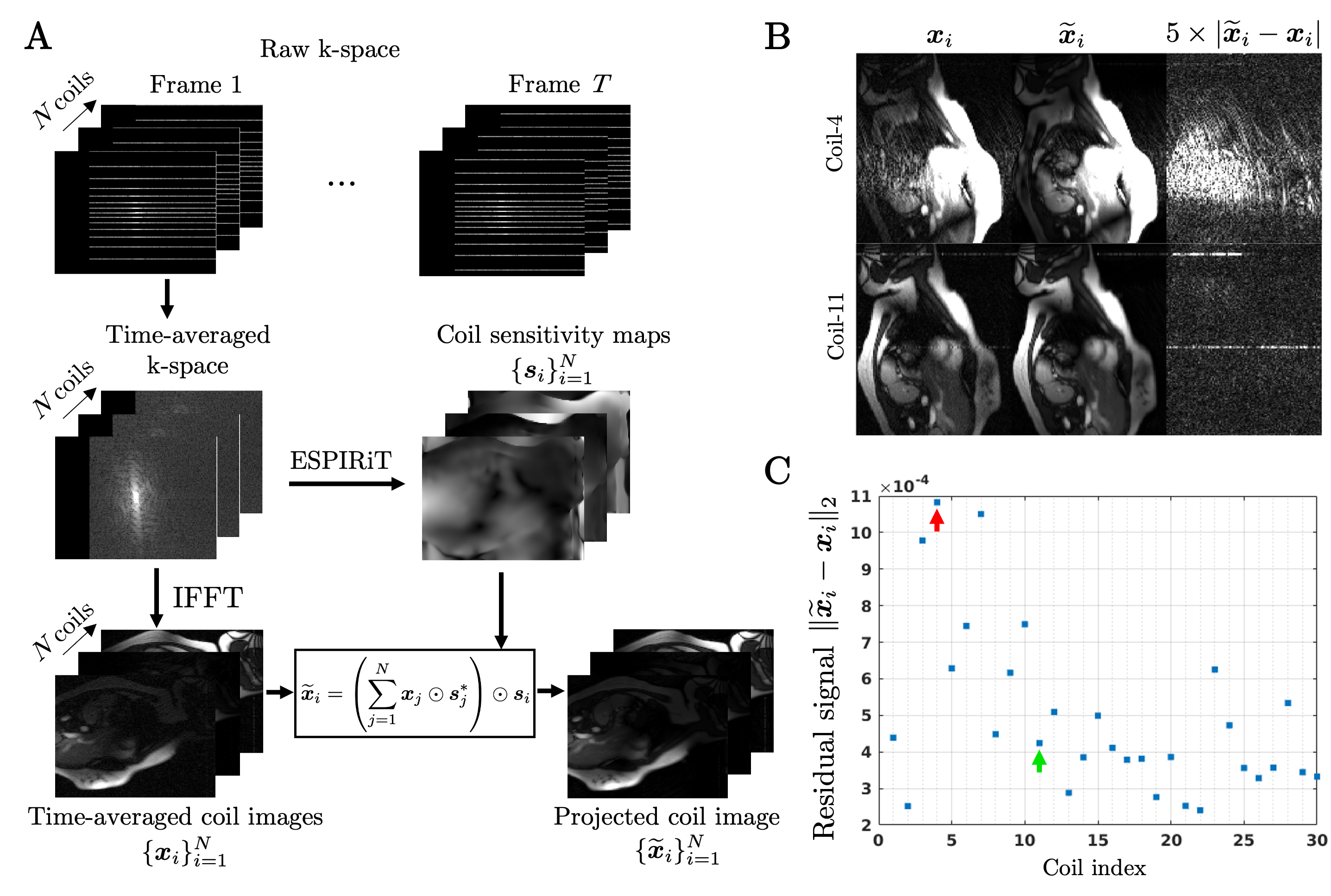}
    \caption{(A) Pipeline to generate time-averaged coil images $\{\vec{x}_i\}_{i=1}^N$ and their re-estimated versions, $\{\tilde{\vec{x}}_i\}_{i=1}^N$. (B) $\vec{x}_i$, $\tvec{x} _i$, and their pixel-wise difference from two coils, one with significant motion artifacts (Coil-4) and one without (Coil-11), are shown. The horizontal lines in Coil-11 are likely from a faint RF interference. This dataset was collected during in-magnet exercise. (C) The residual signal $\|\tvec{x}_i - \vec{x}_i\|_2$ for the $i^{\text{th}}$ coil, with the values from Coil-4 and Coil-11 highlighted with red and green arrows, respectively. For reference, $\|\tvec{x}_i - \vec{x}_i\|_2$ values of all coils were below $6\times 10^{-4}$ at rest (not displayed). }
    \label{fig:reweighting}
\end{figure}

To address this issue, we propose the pipeline shown in \figref{reweighting}A to automatically suppress the contribution from coils that introduce most of the artifacts. We first average the k-space of all the frames to generate the time-averaged k-space for each of the $N$ coils. The coil sensitivity maps $\{\vec{s}_i\}_{i=1}^N$ are then estimated using ESPIRiT \cite{uecker2014espirit}, and the time-averaged coil images $\{\vec{x}_i\}_{i=1}^N$ are derived by performing inverse Fourier transformation of the time-averaged k-space. Subsequently, the time-averaged coil images are first combined and then re-estimated using the coil sensitivity maps, i.e.,
\begin{align}
    \tvec{x}_i = \left(\sum_{j=1}^N \vec{x}_j \odot ~\vec{s}_j^*\right)\odot \vec{s}_i,~\text{for~} i=1,2,\dots,N,
    \label{eq:cr}
\end{align}
where, $\tvec{x}_i$ is the re-estimated time-averaged image from the $i^{\text{th}}$ coil, and ``$\odot$'' and ``$*$'' represent element-wise multiplication and conjugation, respectively. 

Considering that the ESPIRiT coil sensitivity maps are normalized, i.e., $\sum_{j=1}^N \vec{s}_j \odot \vec{s}_j^*$ returns an array of all-ones, one should expect $\tvec{x}_i = \vec{x}_i$ in \eqref{cr} under the ideal condition of $\vec{x}_i = \bvec{x} \odot \vec{s}_i$, where $\bvec{x}$ represents the true coil-combined image. We propose using the residual signal, $\|\tvec{x}_i - \vec{x}_i \|_2$, to identify coil elements that have inconsistent data due to their motion. After computing $\|\tvec{x}_i - \vec{x}_i \|_2$ for each coil, we select $M$ coils with the largest $\|\tvec{x}_i - \vec{x}_i \|_2$ values. The signal for each of those $M$ coils is scaled (reweighted) with $C^2/\|\tvec{x}_i - \vec{x}_i \|_2^2$, where $C$ represents the baseline value computed by averaging residuals from $K$ coils with the smallest residual values. The weighting for the remaining $N-M$ coils is left unchanged. In this work, we chose $M=10$ and $K=15$. For a RT cine image series collected during exercise, \figref{reweighting}B shows the time-averaged and re-estimated images from two physical coils, one with significant motion artifacts (Coil-4) and the other without (Coil-11). The difference between $\vec{x}_i$  and $\tvec{x}_i$ can be mainly attributed to motion artifacts. \figref{reweighting}C shows $\|\tvec{x}_i - \vec{x}_i \|_2$ for all physical coils.



\subsection{Image reconstruction}

The coil reweighting was implemented inline as a pre-processing step and applied to all RT cine data. After coil reweighting, we compressed the k-space data to $12$ virtual coils for faster processing. The image reconstruction for RT cine and flow was also performed inline with Gadgetron-based \cite{hansen2013gadgetron} implementation of a CS method, called sparsity adaptive composite recovery (SCoRe) \cite{chen2019sparsity}. A key feature of SCoRe is that it automatically adjusts the regularization strengths across multiple sparsifying transforms based on the image content. The data-driven and parameter-free nature of SCoRe makes it well-suited for ExCMR, where temporal sparsity can change significantly with the level of exertion. The coil sensitivity maps of the 12 virtual coils were estimated using ESPIRiT \cite{uecker2014espirit}. In the case of RT cine, three-dimensional undecimated wavelet transform (UWT) was applied to sparsify both the spatial and temporal dimensions. For RT flow, in addition to UWT, temporal principal components \cite{petzschner2011fast}, inferred after stacking flow-encoded and flow-compensated images, were used as data-driven sparsifying transform. The reconstruction was performed on a dedicated GPU workstation equipped with NVIDIA GeForce RTX 3090. For a $6$ s per slice acquisition, the reconstruction time was 10 seconds per slice for RT cine and 15 seconds per slice for RT flow.


\subsection{Image analysis}


For cardiac function quantification from RT cine and flow, multiple consecutive heartbeats were available from each slice. 
Several studies have highlighted the impact of respiration-induced variation in cardiac output quantification, which is attributed to changes in intrathoracic pressure  \cite{claessen2014interaction}. 
Also, failing to account for the respiratory motion when selecting cardiac beats from different slices can introduce additional variability due to the through-plane motion. With exercise, this variability is expected to be more pronounced due to exaggerated breathing. To minimize the impact of respiratory motion on cardiac function assessment, heartbeats from the end-expiratory phase were specifically isolated for analysis. 
The respiratory signal, critical for identifying this phase, was derived from RT images using a recently proposed principal component analysis-based method \cite{chen2022ensuring}. 
For imaging at rest, the ECG and respiratory signals facilitated the isolation of a single end-expiratory beat. 
Since the ECG signal was unreliable during exercise, the extracted respiratory signal was superimposed and displayed at the bottom of the exercise RT image series. A moving white marker highlighted the respiratory location for each frame. This signal served as a reference to manually identify systolic and diastolic frames from the end-expiratory phase. An illustration is provided in Supporting Movie S1. The steps of isolating the end-expiratory beat (for rest images) and superimposing the respiratory signal (for exercise images) were performed offline. 




The DICOM images from the short-axis cine and flow acquisitions from the ascending aorta (AAo) and main pulmonary artery (MPA) were imported into suiteHEART (NeoSoft, Pewaukee, WI, USA) for hemodynamic and cardiac function quantification. For the RT data collected at rest, only one end-expiratory beat was imported. However, for RT data acquired during exercise stress, the entire cine and flow series, superimposed with the respiratory signal, were imported. Analyses were conducted only on expiratory heartbeats, aided by the reference respiratory signal. For an initial cohort of ten volunteers imaged with an acquisition time of 3 s/slice, every effort was made to conduct the analyses on the end-expiratory beat where possible. The protocol was consequently modified to an acquisition time of 6 s/slice to accommodate slower breathing patterns at rest. Biventricular segmentation was initially performed using suiteHEART software and manually adjusted where necessary. The following parameters were derived for cardiac function analysis: left ventricular (LV) and right ventricular (RV) end-diastolic volume (EDV), end-systolic volume (ESV), stroke volume (SV), and ejection fraction (EF). From the flow images, peak velocity (Vmax), net forward flow (NFF), and SV were measured for both major arteries. To account for the exaggerated through-plane motion, three RT flow slices were acquired from AAo and MPA. From the three slices available, a slice closest to but above the aortic valve was used for AAo flow quantification, and a slice closest to the pulmonic valve but away from the bifurcation was used for MPA flow analysis. The HR at rest and during each exercise stage was extracted from both RT cine and flow images and converted to a percentage of the age-predicted maximal heart rate (APHR). Cardiac output (CO), computed separately from cine and flow images, was calculated as the product of SV (mL/beat) and HR (beats/min). The values of EDV, ESV, SV, NFF, and CO were indexed using body surface area. In repeat imaging at rest and during a $40$ W workload, quantification was performed on a heartbeat from an arbitrary respiratory phase (AP) and on a heartbeat from end-expiration (EE). The repeatability was assessed using concordance correlation coefficient (CCC) and normalized mean absolute error (NMAE). NMAE, inferred from the two repeats, $x_1$ and $x_2$, was calculated as $\mathrm{NMAE} = \frac{100}{L} \times \sum_{i=1}^{L}\left[(|x_{1,i} - x_{2,i}|)/(\frac{1}{2}|x_{1,i}+x_{2,i}|)\right]$, where $L$ is the number of paired measurements.

To analyze the impact of coil reweighting on image artifacts, the RT-cine series reconstructed with and without coil reweighting were visually scored by two cardiologists for the level of artifacts. Ten healthy volunteers were randomly chosen along with all six patients. From each of these 16 subjects, one mid-ventricular short-axis slice and one two-chamber slice were selected at rest and the maximum level of exertion. The cine series, displayed as movies, were evaluated using a five-point scale: 1--Unusable, 2--Severe artifacts obscuring useful information, 3--Moderate artifacts with some loss of information but still diagnostic, 4--Minor artifacts with minimal loss of information, 5--No artifacts. 
Since the flow images exhibited minimal motion artifacts, coil reweighting was not applied to them. The motion artifacts in the bSSFP-based cine images are more pronounced because the inaccurate sensitivity maps cause the signal from the bright subcutaneous fat to alias to the rest of the image.

To statistically assess the impact of coil reweighting, paired t-tests were conducted using a significance level of 0.05. To account for potential non-independence, we adopted a conservative approach by averaging scores across two views and two readers prior to performing the t-tests. Although Likert-type data are ordinal, the t-test was employed under the assumption that the averaged scores approximate interval-level data. Additionally, inter-reader agreement for artifact scores was evaluated using the weighted Kappa statistic.


\section{Results}\label{sec:res}

\subsection{Quantification results from healthy subjects} 
\begin{figure}[ht]
    \centering
    \begin{subfigure}{.49\textwidth}
        \centering
        \begin{tikzpicture}
            \node[anchor=south west,inner sep=0] (image) at (0,0) {
                \includegraphics[width=\linewidth]{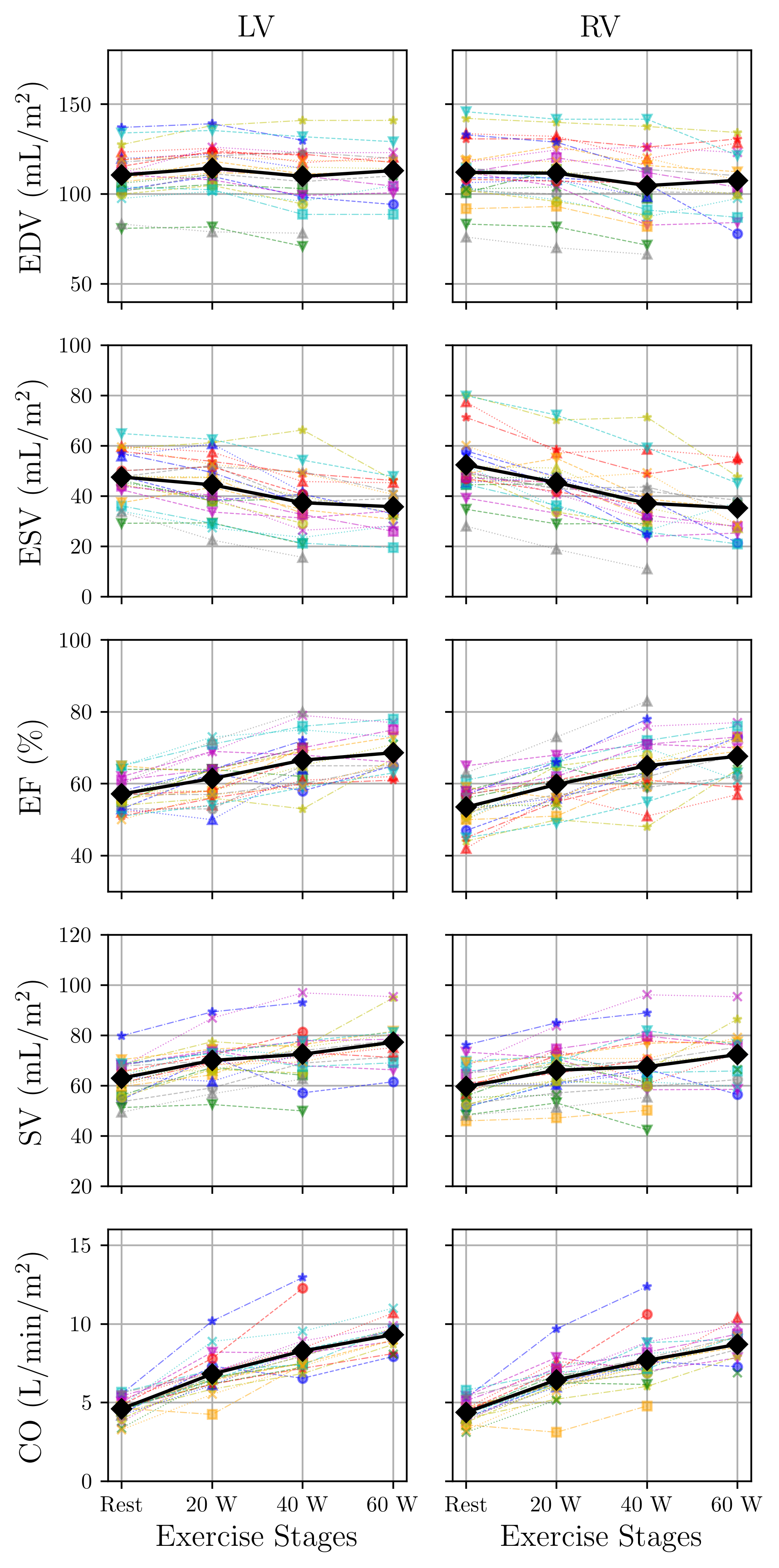}
            };
            \node[anchor=north west, xshift=0pt, yshift=5pt] at (image.north west) {\textbf{A}};
        \end{tikzpicture}
        \label{fig:cine-quant}
    \end{subfigure}\hfill
    \begin{subfigure}{.49\textwidth}
        \centering
        \begin{tikzpicture}
            \node[anchor=south west,inner sep=0] (image) at (0,0) {
                \includegraphics[width=\linewidth]{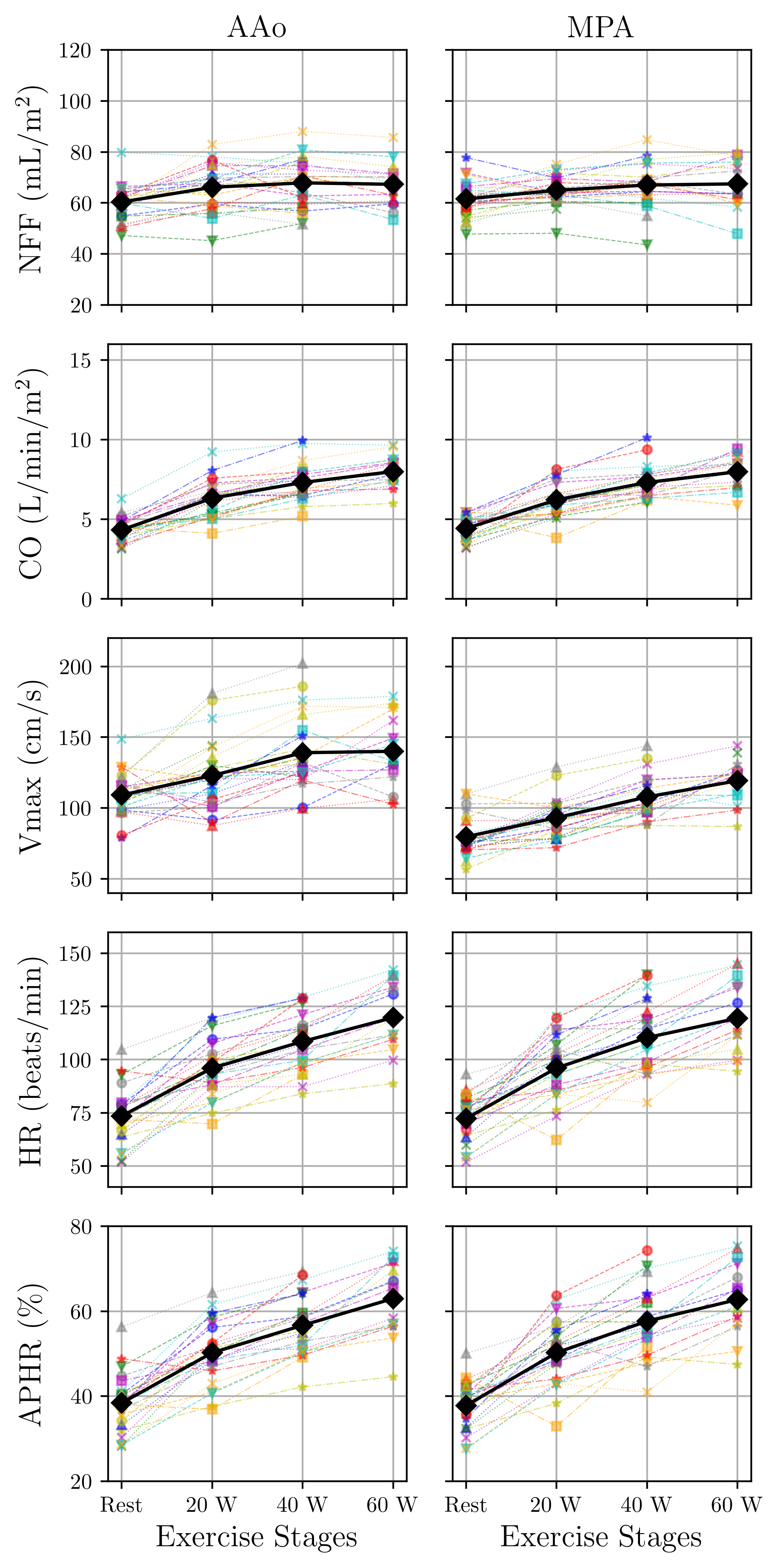}
            };
            \node[anchor=north west, xshift=0pt, yshift=5pt] at (image.north west) {\textbf{B}};
        \end{tikzpicture}
        \label{fig:flow-quant}
    \end{subfigure}
    \caption{Biventricular cardiac function and hemodynamic assessment from RT cine (A) and RT flow (B). Each trace represents a volunteer, and the thicker black line represents the average value. The HR and APHR values from cine (not shown) were similar to that from flow.}
    \label{fig:cine-flow-quant}
\end{figure}

The image quality, after coil reweighting, was deemed adequate to perform quantification in all cases. 
In \figref{cine-flow-quant}A, the cardiac function parameters extracted from the short-axis RT cine images are displayed, tracing the progression from rest to three stages of exercise for each volunteer. In summary, the EDV remained relatively consistent from rest through the exercise stages in both ventricles. In contrast, the ESV decreased as exercise intensity and HR increased, indicating more vigorous cardiac contraction during exercise. Consequently, the SV typically increased under stress, which is consistent with the literature \cite{beaudry2018exercise}. However, the pattern of increase in SV varied among volunteers, ranging from a rapid initial rise to plateauing in some, and a more gradual increase in others. The consistent elevation in CO was attributable to the increased SV and HR. The HR and APHR trends extracted from the cine images (not shown) were similar to the ones extracted from the flow images discussed below. 

In \figref{cine-flow-quant}B, the aortic and pulmonic flow parameters, tracked from rest through three stages of exercise for each volunteer, are depicted. The NFF values demonstrate an overall increase with exercise, although this rise is less pronounced compared to the biventricular stroke volumes. Nonetheless, CO exhibits a similar trend of increasing with exercise intensity. The peak velocities for both arteries also show an upward trend with exercise. 
Across all volunteers, the maximum heart rates (recorded from cine or flow) ranged from 92 to 152 beats per minute (bpm) and the maximum APHR ranged from 46\% to 81\%. On average, the subjects were able to achieve a maximum APHR value of 70\%. The perceived exertion, measured using BRPE, ranged from 9 to 16, with an average value of 12.3. For the subjects who quit before reaching the $60$ W workload, the most commonly expressed reasons were shortness of breath, leg fatigue, and overall exhaustion. 

\begin{figure}
    \centering
    \includegraphics[width=\linewidth]{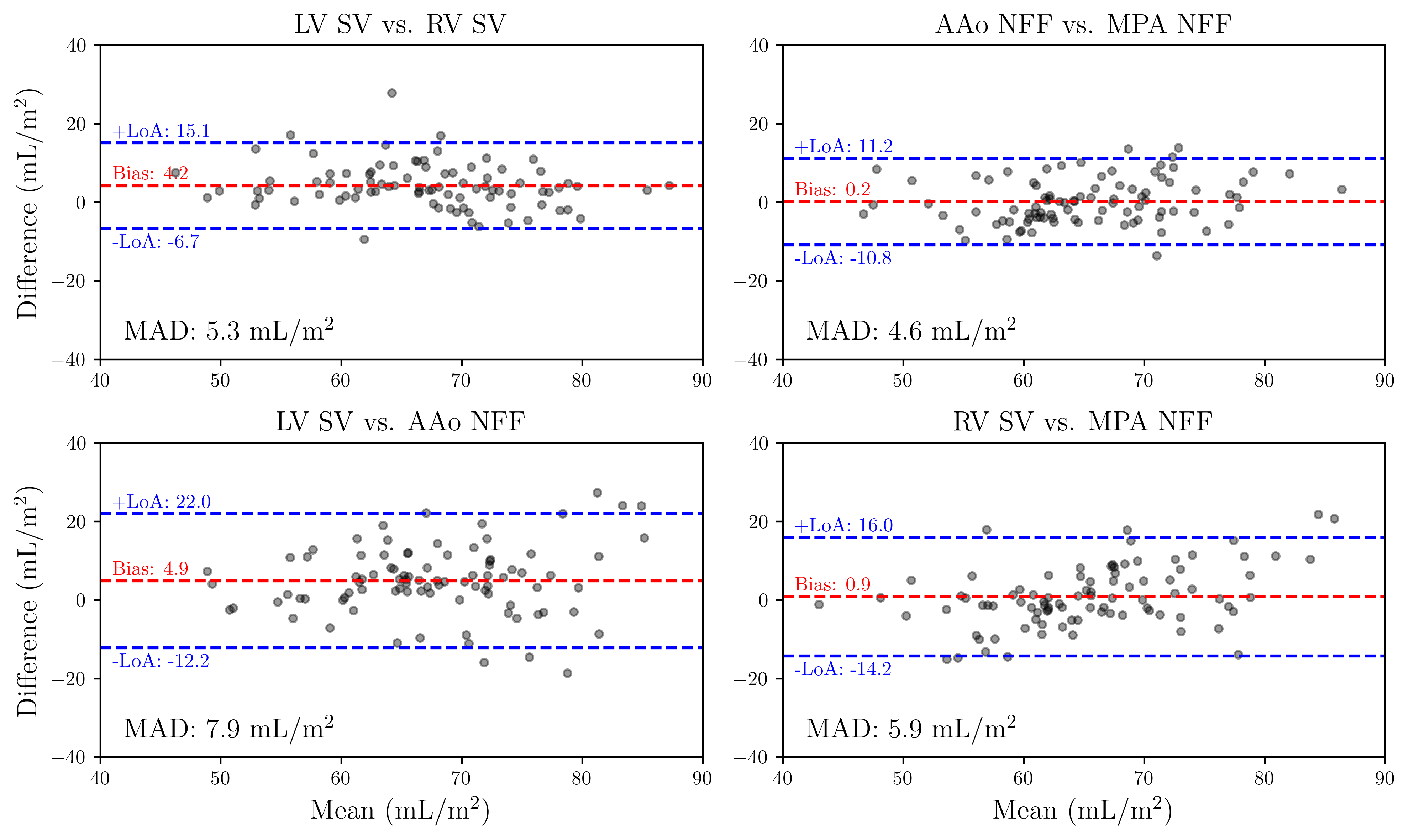}
    \caption{Bland-Altman plots demonstrating the consistency of LV and RV stroke volumes from cine and flow imaging. This analysis includes data pooled from rest and all exercise stages. Top-left: LV and RV SV from cine. Top-right: AAo NFF and MPA NFF from flow imaging. Bottom-left: LV SV from cine and AAo NFF from flow imaging. Bottom-right: RV SV from cine and MPA NFF from flow imaging.}
    \label{fig:cine-flow-ba}
\end{figure}

The Bland-Altman analysis demonstrating the consistency of LV and RV stroke volumes inferred from cine and flow imaging is shown in \figref{cine-flow-ba}. This analysis includes data pooled from rest and all exercise stages. Cine-based LV vs. RV SV exhibits a small bias (4.2 mL/m$^2$) and a low mean absolute difference (MAD) (5.3 mL/m$^2$), while aortic vs. pulmonic NFF shows an even lower bias (0.2 mL/m$^2$) and MAD (4.6 mL/m$^2$) values. Comparisons between cine- and flow-derived volumes show slightly higher discrepancies, with LV SV vs. AAo NFF yielding a bias of 4.9 mL/m$^2$ and  MAD of 7.9 mL/m$^2$, and RV SV vs. MPA NFF showing a bias of 0.9 mL/m$^2$ and a MAD of 5.9 mL/m$^2$. The weaker agreement between cine and flow may be attributed to: (i) differences in imaging parameters, (ii) mild mitral and tricuspid regurgitation commonly observed in healthy subjects \cite{klein1990age}, (iii) coronary flow, (iv) inaccuracies in background phase correction in flow imaging, (v) inclusion of trabeculation in cine segmentation \cite{han2009impact}, which may represent a larger part of ESV during exercise, and (vi) heart rate variations between cine and flow. Nonetheless, the bias and MAD remain below 5 mL/m$^2$ and 8 mL/m$^2$, respectively, across all four comparisons, indicating overall consistency.

\begin{figure}[ht]
    \centering
    \includegraphics[width=\textwidth]{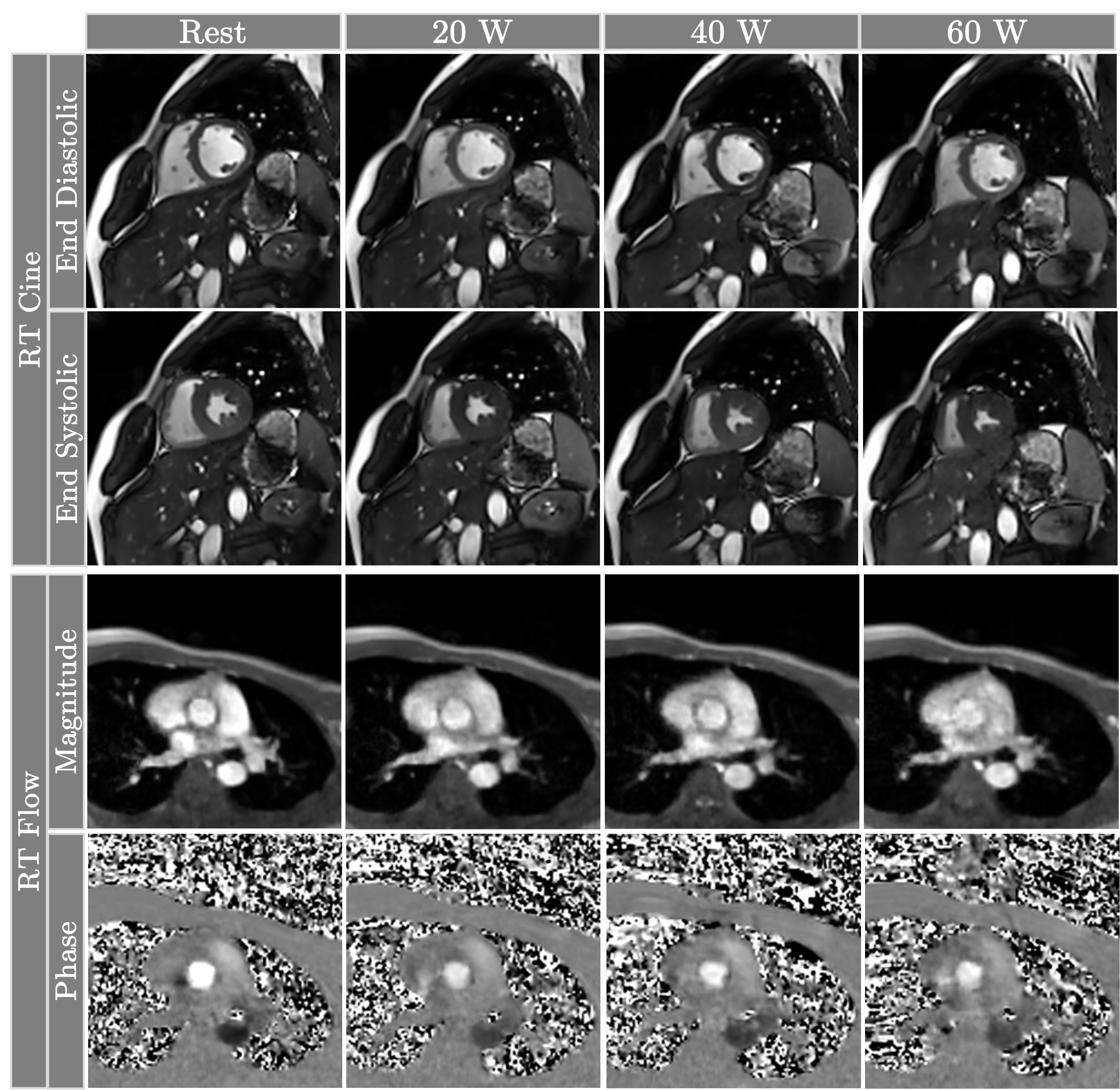}
    \caption{Representative cine and flow frames from a healthy volunteer at rest and across three stages of exercise. Stronger contractility under exercise stress is observed from end-systolic frames (second row). The brighter phase (bottom row) in the resting image is due to the lower value of VENC.}
    \label{fig:healthy-frame}
\end{figure}

Representative cine and flow images at rest and various stages of exercise from a healthy volunteer are shown in \figref{healthy-frame}. The cine images show a clear delineation of the blood-myocardium boundary, with some marginal blurring at $60$ W due to the residual motion. The end-systolic frames exhibit stronger contractility, especially at $60$ W. The flow images are mostly free of motion artifacts, with the $60$ W magnitude images showing marginal blurring. Note that a higher value of velocity encoding gradient (VENC) was used during exercise; therefore, the velocities are scaled differently for images collected during exercise compared to the resting images.

\begin{figure}[ht]
    \centering
    \includegraphics[width=\textwidth]{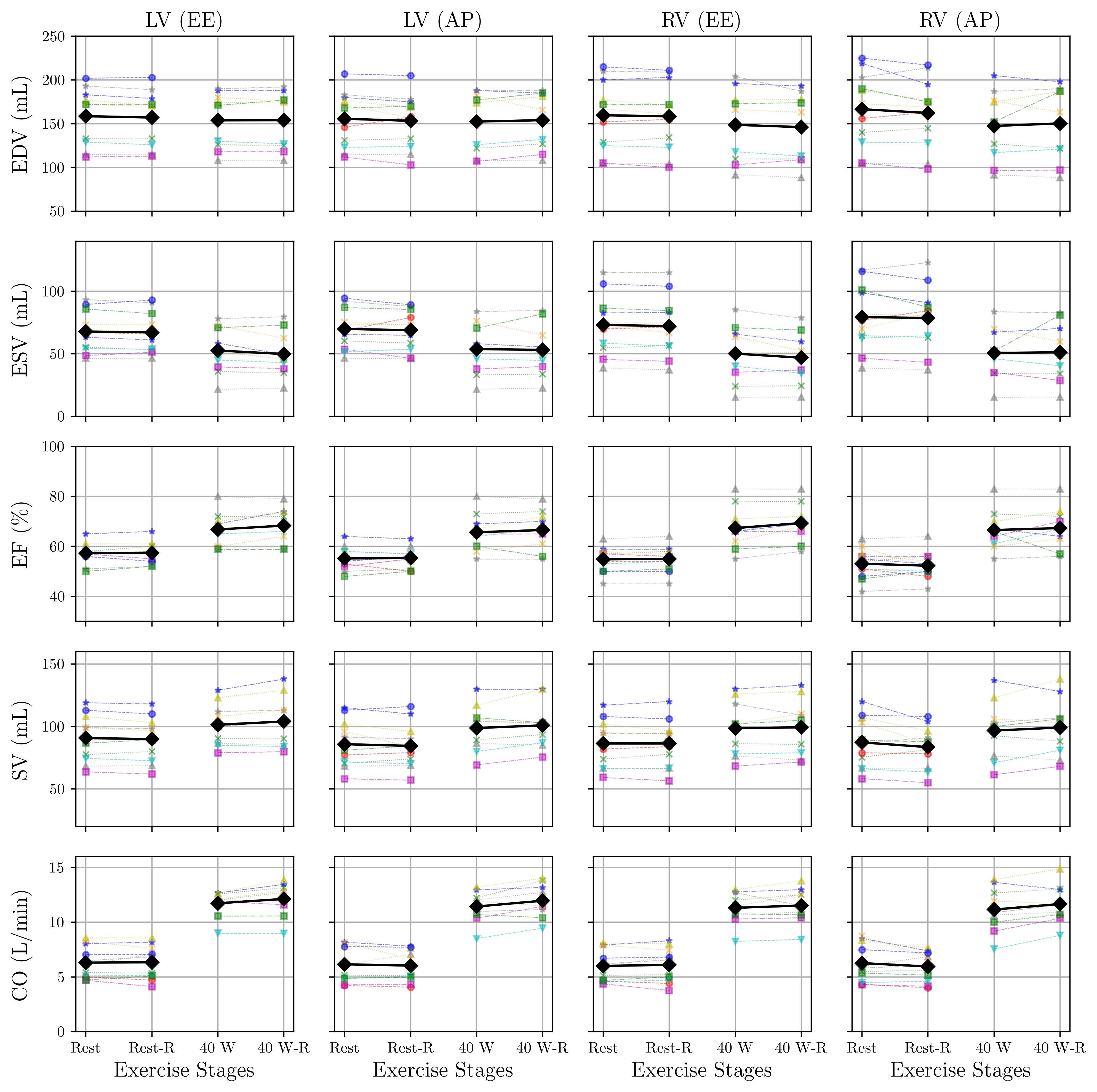}
    \caption{Repeatability of the cine parameters at rest and $40$ W for beats from arbitrary respiratory phase (AP) and beats from end-expiratory phase (EE). Here, ``Rest-R'' and ``$40$ W-R'' represent the repeated measurements.}
    \label{fig:cine-rep}
\end{figure}

\begin{figure}[ht]
    \centering
    \includegraphics[width=\textwidth]{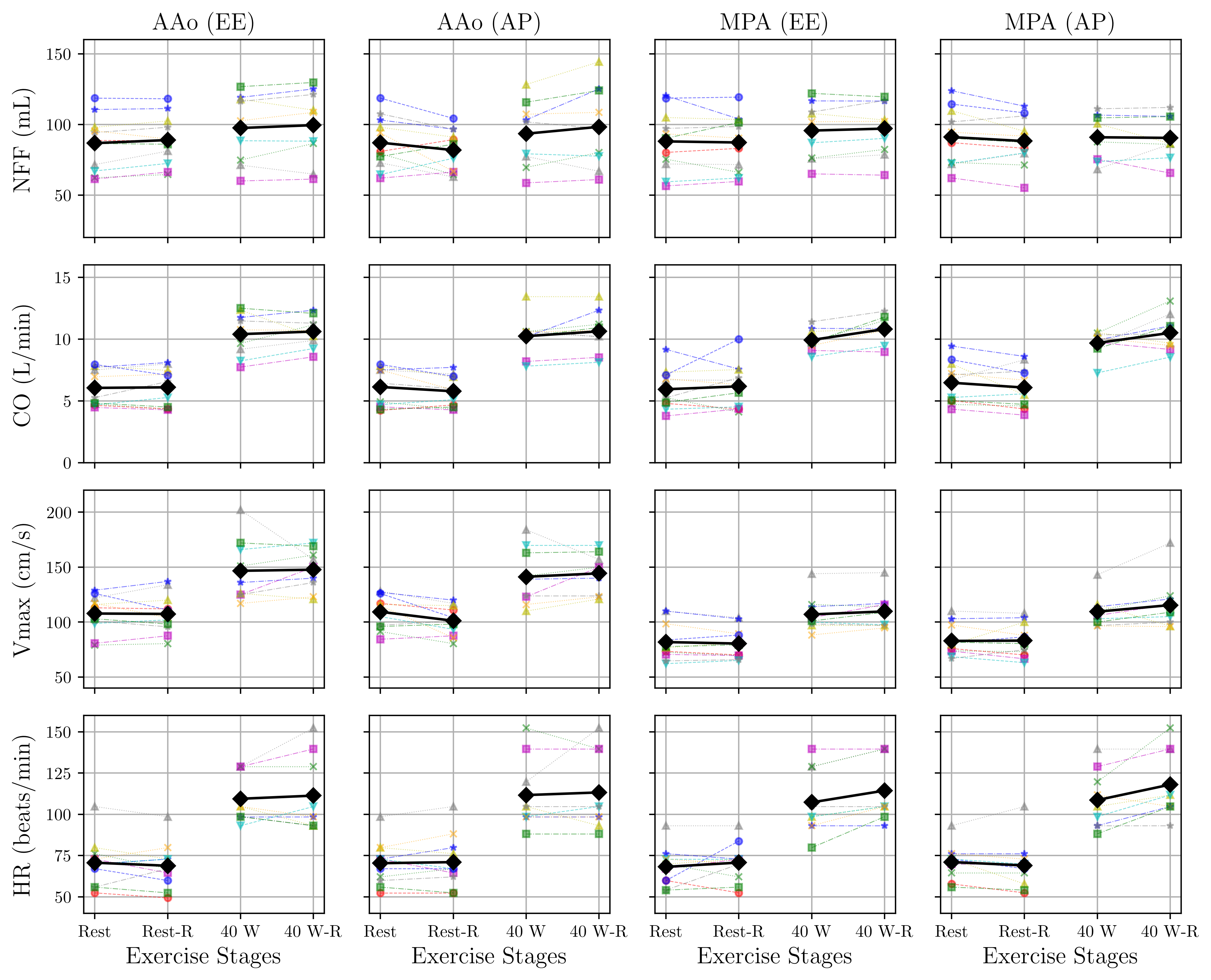}
    \caption{Repeatability of the flow parameters at rest and $40$ W for beats from arbitrary respiratory phase (AP) and beats from end-expiratory phase (EE). Here, ``Rest-R'' and ``$40$ W-R'' represent the repeated measurements.}
    \label{fig:flow-rep}
\end{figure}

\begin{figure}[ht]
    \centering
    \includegraphics[width=\textwidth]{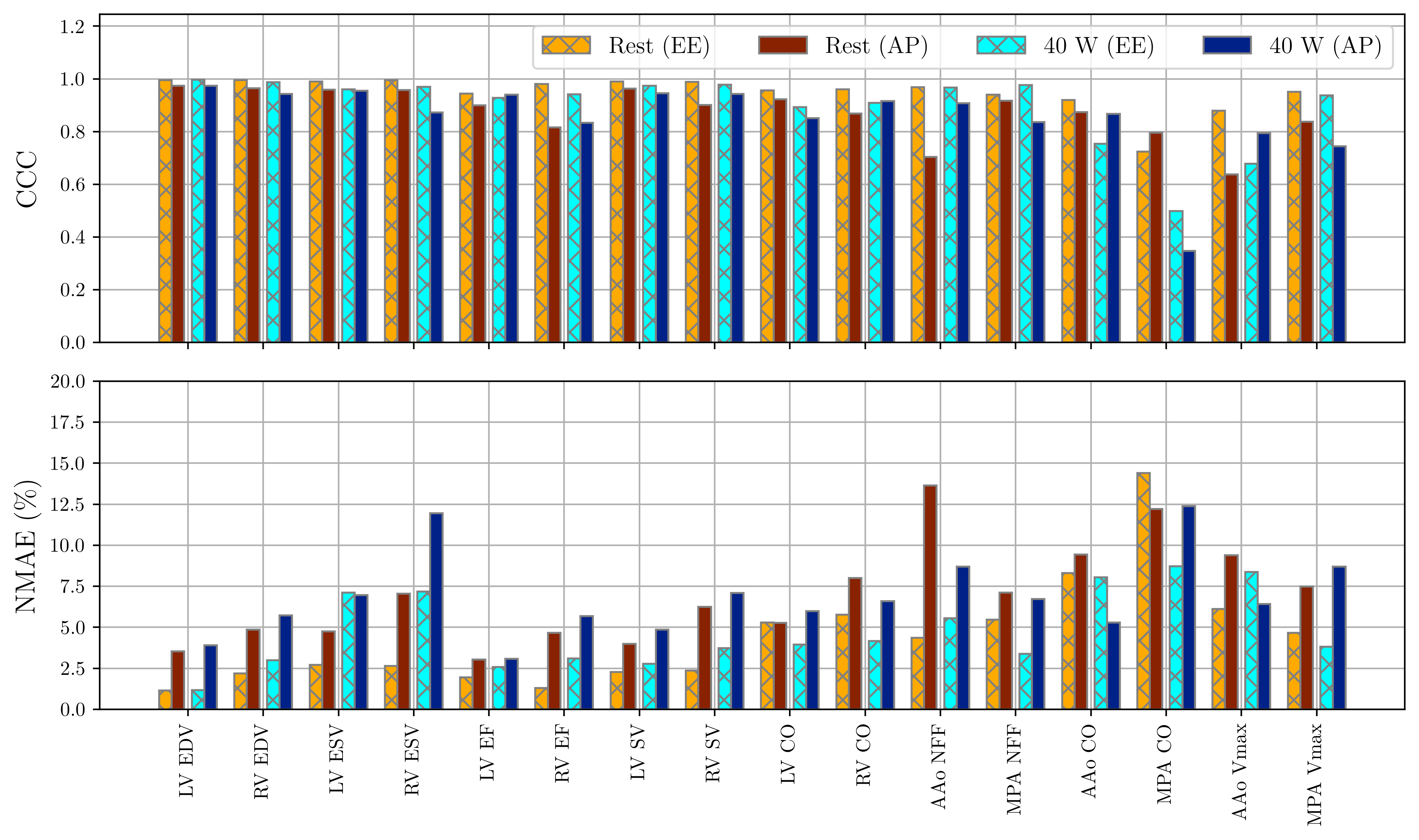}
    \caption{Concordance correlation coefficient (CCC) and normalized mean absolute error (NMAE) for cine and flow parameters for arbitrary respiratory phase (AP) and end-expiratory phase (EE) beats at rest and $40$ W.}
    \label{fig:ccc}
\end{figure}

\figref{cine-rep} displays the repeat assessment of biventricular functional parameters from RT cine images at rest and the workload of $40$ W. Similarly, \figref{flow-rep} presents the repeat assessment of hemodynamic parameters from RT flow images at the same conditions. Overall, both AP and EE beats demonstrate a high level of repeatability. However, as shown in \figref{ccc}, in terms of CCC and NMAE, EE beats show a marginal but consistent improvement over AP beats. With the sole exception of MPA CO, NMAE from all parameters from the EE beat is less than 10\%. In contrast, the NMAE exceeds the 10\% threshold four times for AP. A similar trend is observed in CCC. 


\subsection{Quantification results from patients}
\begin{figure}[ht]
    \centering
    \includegraphics[width=\textwidth]{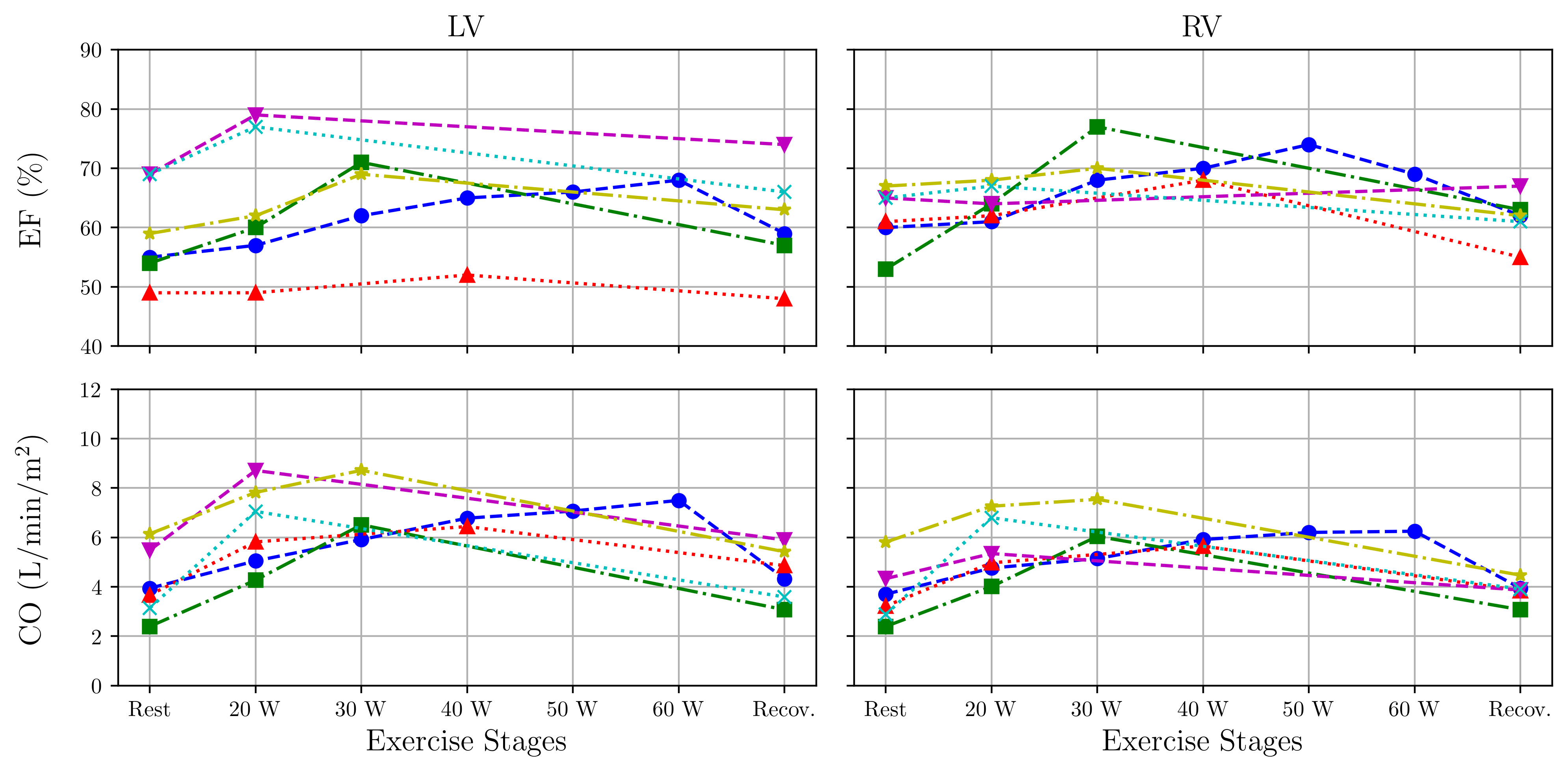}
    \caption{Biventricular cardiac function assessment from RT cine acquired from six patients. The purple trace represents a patient where inadequate LV coverage due to user error led to a discrepancy between RV and LV quantification. Here, ``Recov.'' represents post-exercise recovery.}
    \label{fig:patient-quant}
\end{figure}

The RT cine data were reconstructed from six patients. 
In all cases, the image quality, after coil reweighting, was deemed adequate for performing cardiac function. \figref{patient-quant} shows biventricular EF and CO values at rest, during exercise, and post-exercise recovery. Due to the differences in the workload increments and widely different exercise capacities, it was not possible to average the data across patients. Except for one patient, where inadequate spatial coverage due to a user error led to LV-RV mismatch, there was a reasonable agreement between CO and EF from LV and RV. Across all patients, the maximum heart rates (recorded from cine) ranged from 91 to 123 bpm and the maximum APHR ranged from 62\% to 81\%. On average, the patients were able to achieve a maximum APHR value of 69\%. The perceived exertion ranged from $13$ to $19$ on the BRPE scale, with an average value of $16.2$. 

\begin{figure}[ht]
    \centering
    \includegraphics[width=\textwidth]{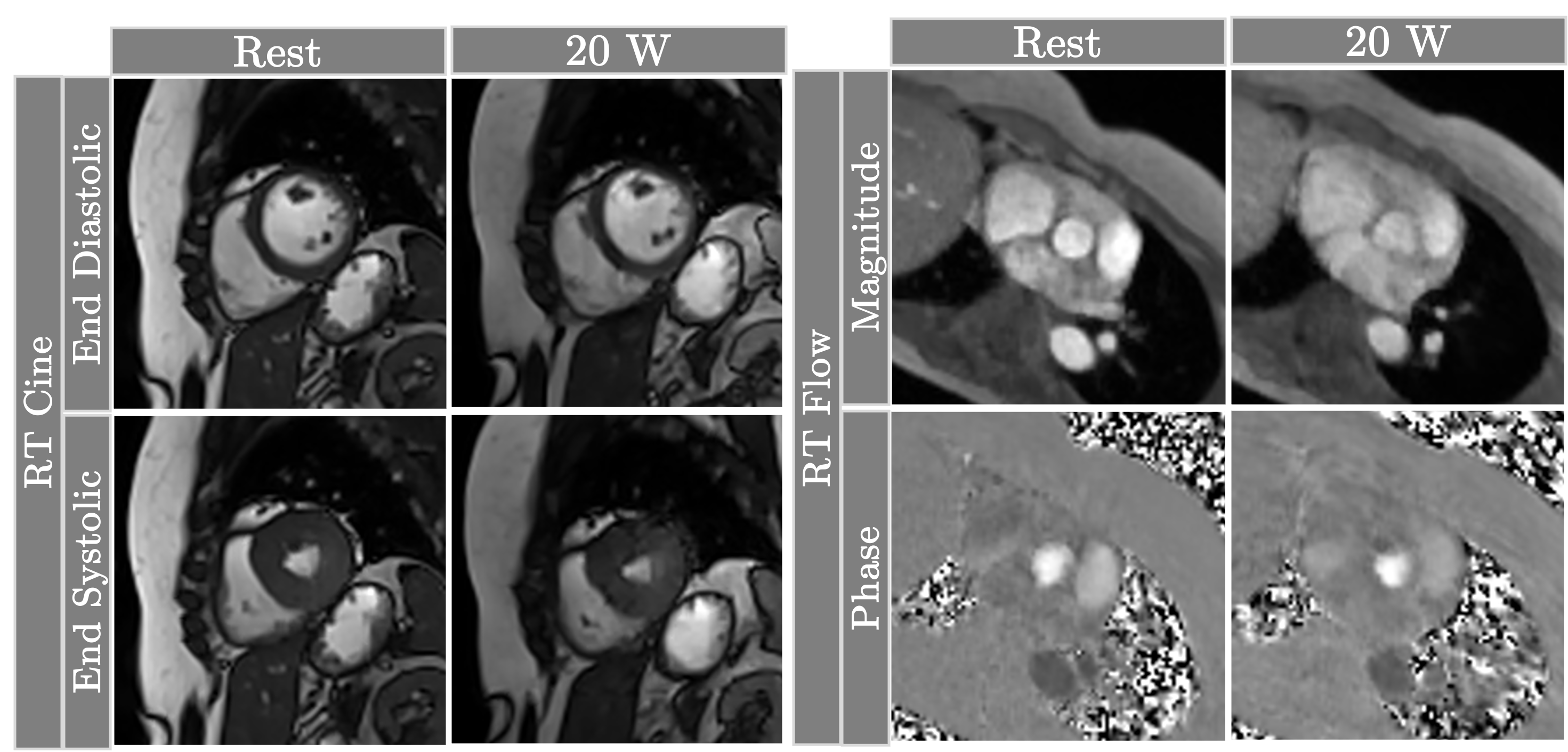}
    \caption{Representative cine and flow frames from a patient at rest and $20$ W exercise. Stronger contractility under exercise stress is observed from end-systolic frames. The brighter phase in the resting image is due to the lower value of VENC.}
    \label{fig:patient-frame}
\end{figure}

 \figref{patient-frame} shows a representative frame from a patient fatigued at the workload of $20$ W. Both RT cine images show clear delineation of blood and myocardium during the systolic and diastolic phases. The flow images are also mostly free of artifacts, with the $20$ W magnitude image showing slight blurring. The phase images, however, clearly capture the flow in the ascending and descending aorta. Note that a higher value of VENC was used at $20$ W; therefore, the velocities are scaled differently in the $20$ W image compared to the resting image. Although less comprehensive compared to the healthy volunteer study, these results indicate the feasibility of implementing ExCMR in a clinical setting.

\subsection{Impact of coil reweighting}
\begin{figure}[ht]
    \centering
    \includegraphics[width=\textwidth]{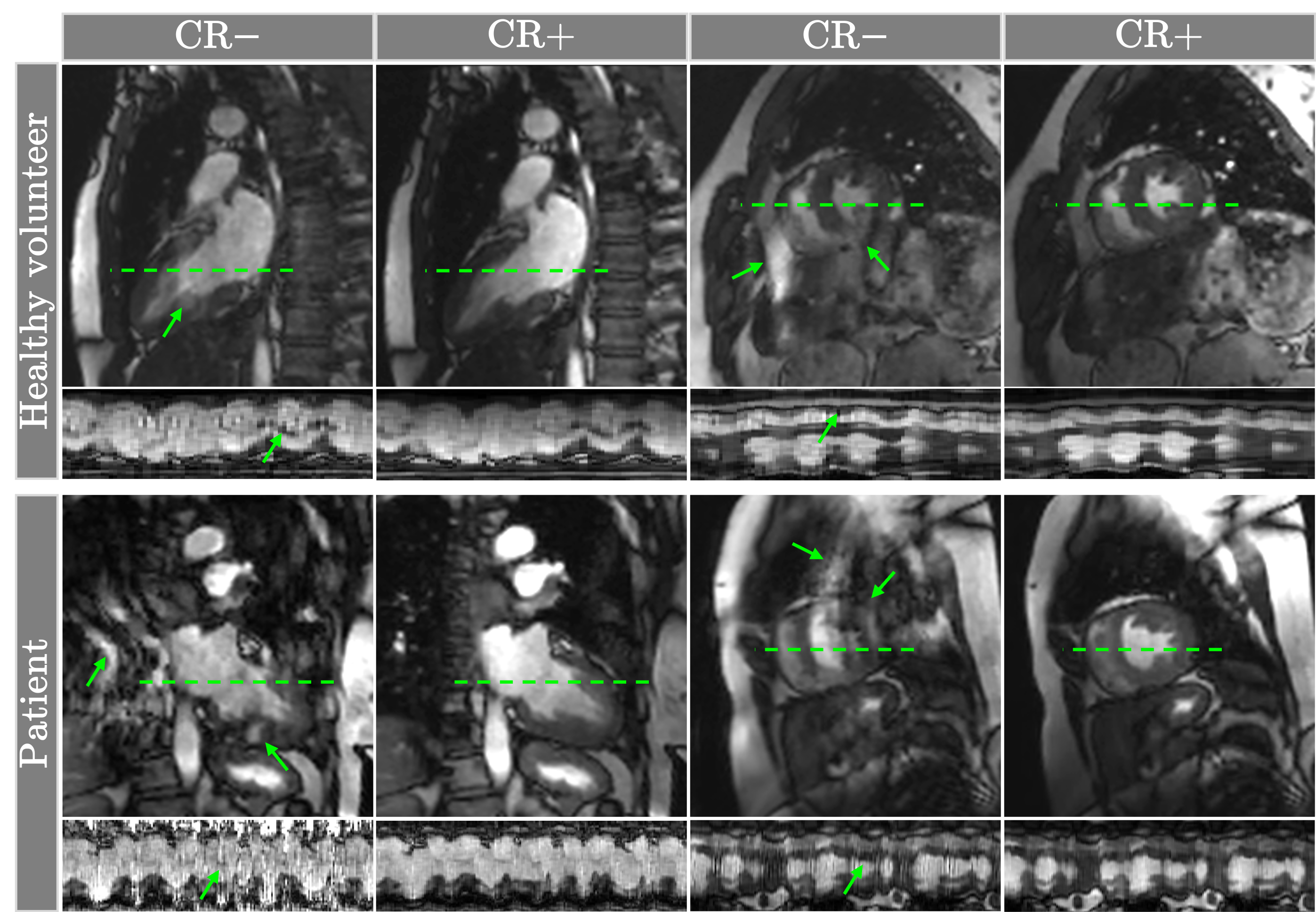}
    \caption{Exercise images reconstructed without (CR$-$) and with (CR$+$) coil reweighting. Temporal profiles along the green dashed lines are shown, with green arrows highlighting the motion artifacts. The top row displays two views from a healthy volunteer, and the bottom row shows two views from a patient.}
    \label{fig:CR-frame}
\end{figure}

The artifacts scores are reported in \tabref{scores}. The difference was not significant at rest ($p=0.14$) but was significant ($p<0.01$) under exercise. 
The inter-reader agreement, as assessed using the weighted Kappa statistic, was 0.653, indicating substantial agreement between the two readers. 
\figref{CR-frame} shows examples from a healthy subject and a patient, where images without coil reweighting exhibit severe artifacts; these artifacts are mostly mitigated after coil reweighting.


\begin{table}[h!]
\centering
\begin{tabular}{|r|c|c|c|c|}
\hline
 & \textbf{Rest CR$-$} & \textbf{Rest CR$+$} & \textbf{Stress CR$-$} & \textbf{Stress CR$+$} \\
\hline
\hline
Artifacts ($n=16$) & 4.81 & \textbf{4.91} & 3.61 & \textbf{4.58} \\
\hline
\end{tabular}
\caption{Artifact levels from ten healthy subjects and six patients before (CR$-$) and after (CR$+$) coil reweighting. Each reported number represents an average over subjects, two views (short-axis and two-chamber), and two readers.}
\label{tab:scores}
\end{table}

\section{Discussion}\label{sec:dis}
ExCMR with in-magnet exercise allows dynamic monitoring of cardiac function during exercise, revealing functional impairments that are not apparent during resting cardiovascular exams. Because breath-holding is generally not feasible during exercise, free-breathing RT imaging is often used to collect data. Technical challenges associated with imaging during exercise include image quality degradation due to motion artifacts and limited spatial and temporal resolution. In this work, we demonstrate the feasibility of in-magnet ExCMR for measuring biventricular cardiac function and hemodynamics during staged exercise in both healthy subjects and patients. 

To ensure adequate spatial and temporal resolutions, we employed high acceleration rates. This was facilitated by variable density Cartesian sampling and parameter-free CS-based inline reconstruction. To further enhance image quality, we proposed and applied a coil reweighting technique that effectively minimizes motion artifacts caused by the physical movement of the MRI receive coils. Additionally, we leveraged our newly proposed technique to extract respiratory signal directly from the imaging data. This technique allows for consistent selection of end-expiratory heartbeats, thereby minimizing the impact of through-plane motion and improving the repeatability of imaging parameters.  In summary, we are able to address the aforementioned limitations for in-magnet ExCMR by (i) achieving high acceleration rates of $8-9$ in RT cine and $16-19$ in RT flow, leading to the spatial resolution of $1.7-2.4$ mm for cine and $2-2.5$ mm for flow, (ii) reducing scan time such that a full short-axis stack can be acquired in less than two minutes, (iii) providing whole heart coverage, (iv) developing data processing that is independent of ECG, (v) providing biventricular quantification, (vi) yielding hemodynamic assessment, (vii) developing inline image reconstruction, (viii) establishing scan-rescan repeatability, (ix) suppressing motion artifacts by coil reweighting, (x) enabling images across multiple exercise intensities, and (xi) mitigating sensitivity to through-plane motion by using end-expiratory beats for quantification.



A key component of our framework, coil reweighting, suppressed motion artifacts and enabled quantification from highly accelerated RT imaging across varying exercise intensities. Without coil reweighting, some of the cine images were not interpretable or suitable for quantification. With coil reweighting, most of the artifacts were eliminated. Equally importantly, coil reweighting did not negatively impact images collected at rest. Compared to coil rejection, which can result in a more significant signal loss in certain image areas, coil reweighting offers a softer approach. Nevertheless, coil reweighting suppresses contributions from certain coil elements, which can lead to partial loss of image intensity. However, we did not observe a case where the benefit of suppressing artifacts was outweighed by the loss of intensity.

The performance of the proposed coil reweighting approach depends on the user-defined values of $K$ and $M$. The parameter $K$ represents the number of coil elements that do not experience significant movement. In our MRI experiments, there were $K_{\sf b}=18$ body coil elements and $K_{\sf s}=12$ supine coil elements. Since the supine coil elements are fixed in position and only a fraction of body coil elements have significant movement, we found that setting $K$ equal to or slightly larger than $K_{\sf s}$ is a reasonable choice. In our experience, selecting a value in the range $K_{\sf s} \leq K \leq 20$ gives comparable results. The other parameter $M$ can be considered as an upper limit on the number of coil elements that experience significant movement. When we tracked the change in $\|\tvec{x}_i - \vec{x}_i\|_2$ from clean images at rest to degraded images under exercise, we observed that typically 2 to 8 coil elements had a marked increase in $\|\tvec{x}_i - \vec{x}_i\|_2$. We conservatively selected $M=10$ based on this observation. In our experience, $8\leq M \leq K_{\sf b}$ gives comparable results.


The limitations of this work include a small sample size, primarily comprising healthy subjects with low average BMI. All studies were conducted on a 3T scanner, and extending this work to lower field strengths may result in reduced image quality. The bore size and length of the magnet also pose challenges; some larger patients may not be able to exercise comfortably inside the magnet. Additionally, this particular protocol had an extended exercise duration, primarily driven by the inclusion of 4D flow acquisition at each exercise stage. If limited to RT cine and flow measurements, it is feasible to complete scanning at each stage within three minutes and dramatically shorten the exercise protocol to 15 minutes for a 5-staged exercise protocol similar to the Bruce protocol \cite{bruce1973maximal}. Furthermore, the impact of coil reweighting on cardiac function quantification for imaging at rest, where a breath-held reference is available, remains to be validated. Moreover, although designed for a single set of sensitivity maps, our coil-reweighting method could be extended to multiple sets by incorporating set-specific maps into \eqref{cr} and summing over them. However, its effectiveness with multiple sets of sensitivity maps has yet to be validated. Another limitation is that the highest achieved heart rate in the volunteers and patients was 152 bpm and 123 bpm, respectively. Future research directions include further optimizing the ExCMR protocol to achieve higher levels of exertion, improving temporal resolution to enable RT imaging at maximal exercise, and extending coil reweighting to 3D applications.

\section{Conclusion}\label{sec:con}
In this preliminary study, we demonstrated the feasibility of ExCMR with in-magnet exercise for quantitative cardiac function evaluation in both healthy volunteers and patients under multi-stage exercise. The highly accelerated RT imaging was facilitated by a parameter-free CS, overcoming the traditional challenges of limited resolution. Additional key innovative aspects of our technique are the incorporation of coil reweighting and the selection of end-expiratory heartbeats, which resulted in significantly improved image quality and enhanced repeatability of cardiac function measurements. 

\clearpage

\clearpage

\clearpage

\clearpage

\clearpage

\clearpage

\clearpage

\clearpage

\clearpage

\clearpage

\clearpage

\clearpage

\clearpage


\begin{figure}[h]
    \centering
    \caption*{{\bf{Supporting Movie S1:}} The movie shows a 6 s cine image series from a healthy volunteer collected under $60$ W of exercise. The curve at the bottom represents the respiratory motion, with the moving cross highlighting the current frame. }
    \label{fig:resp-curve}
\end{figure}

\clearpage

\backmatter

\bmhead{Acknowledgments and Funding}
This work was funded by NIH grants R01-EB029957, R01-HL151697, R01-HL148103, and R01-HL135489.

\section*{Ethics Declarations}
\begin{itemize}
\item \textbf{Competing interests}: The authors declare no competing interests.
\item \textbf{Ethics approval and consent}: For the human subject data, approval was granted by the Institutional Review Board (IRB) at The Ohio State University (2019H0076). Informed consent to participate in the study and publish results was obtained from all individual participants.
\item \textbf{Data and code availability}: MRI data and the code to generate results are available upon request
\item \textbf{Author contribution}: P. Chandrasekaran assisted with data acquisition and processing as well as manuscript writing, C. Chen implemented data processing techniques, Y. Liu assisted with data acquisition and pulse sequence programming, S.M. Arshad assisted with experiment planning, C. Crabtree assisted with subject monitoring during exercise, M. Tong and Y. Han assisted with experiment design and results interpretation, and R. Ahmad supervised all aspects of the study.
\end{itemize}

\bibliography{references}
\end{document}